\documentclass{article}
\usepackage[T1]{fontenc} 
\usepackage[utf8]{inputenc} 
\usepackage{ismir,amsmath,cite,url}
\usepackage{graphicx}
\usepackage{color}
\usepackage{footnote}
\makesavenoteenv{tabular}

\usepackage{siunitx} 
\usepackage{booktabs} 
\usepackage[bookmarks=false,hidelinks]{hyperref}

\usepackage{tabularx}

\usepackage{float}
\usepackage{xcolor}

\usepackage{tikz}

\usepackage{xspace}
\usepackage{multirow}

\newcommand{\datasetname}{\textit{Batik-plays-Mozart}\xspace}
\newcommand{\repo}{\url{https://github.com/huispaty/batik\_plays\_mozart}\xspace} 

\usepackage{lineno}

\title{The Batik-plays-Mozart Corpus: 
Linking \\Performance to Score to Musicological Annotations}

\twoauthors
 {Patricia Hu} 
 {Gerhard Widmer} 
 {Institute of Computational Perception, Johannes Kepler University Linz, Austria} 
 {LIT AI Lab, Linz Institute of Technology, Austria} 
 {1} 
 {2} 
 {\texttt{firstname.lastname@jku.at}} 
    
\def\authorname{P. Hu and G. Widmer}

\begin{document}

%
\maketitle
\begin{abstract}
We present the \datasetname Corpus, a piano performance dataset
combining professional Mozart piano sonata performances with expert-labelled scores
at a note-precise level. The performances originate from a recording by Viennese pianist Roland Batik
on a computer-monitored Bösendorfer grand piano, and are available both as MIDI files and
audio recordings. They have been precisely aligned, note by note, with a current standard
edition of the corresponding scores (the New Mozart Edition) in such a way that they can
further be connected to the musicological annotations (harmony, cadences,
phrases) on these scores that were recently published by \cite{hentschel2021annotated}. 

The result is a high-quality, high-precision corpus mapping scores and musical
structure annotations to precise note-level professional performance information.
As the first of its kind, it can serve as a valuable resource for studying various facets
of expressive performance and their relationship with structural aspects.

In the paper, we outline the curation process of the alignment and conduct two
exploratory experiments to demonstrate its usefulness in analyzing expressive performance.

\end{abstract}
\section{Introduction}



Music performance is a complex and nuanced activity that involves the interplay of various expressive features such as timing, dynamics, and articulation. Expressive performance research in music information retrieval (MIR) focuses on modeling expressive aspects of music performance by analyzing how performers use nuances in timing, dynamics, articulation, and other expressive features to convey their musical intentions, with the aim of developing computational models that can analyze, recognize, or synthesize expressive performances \cite{cancino2018computational}.

Recent research in this field for Western classical piano has focused on data-driven approaches both for performance generation \cite{jeong2019virtuosonet, fujishima2018rendering} and data creation in the form of large-scale MIDI performance data transcribed from audio recordings 
\cite{zhang2022atepp, kong2020giantmidi}. While such data corpora can be useful for comparative performance analyses and related tasks (e.g., performer identification, performance style transfer), they lack the necessary precision and alignment information (with the underlying musical score) required to precisely map expressive intentions and parameters to underlying score features.

Compared to these large-scale transcribed MIDI datasets, precise MIDI data (as recorded on computer controlled grand pianos such as the Yamaha Disklavier or Boesendorfer SE/Ceus series) along with their corresponding score alignment is somewhat limited in quantity and size \cite{foscarin2020asap, vienna4x22, hashida2018crestmusepedb}. The performances in such datasets are typically sourced from advanced piano students or piano competitions, whereas the digital scores are often obtained from open-source, user-curated online libraries such as MuseScore\footnote{\url{https://musescore.com/sheetmusic}}.

Regarding the performance-to-score alignment, one would ideally want to have note-by-note correspondence information; unfortunately, in the case of the largest of these datasets \cite{foscarin2020asap}, score-performance alignments are only given at a rather coarse level of beats. Score annotations conveying structural information such as underlying harmony or phrases are even more scarce. 

To address these limitations, we introduce the \datasetname dataset\footnote{\repo}, in which we provide a set of expert performances of 12 complete Mozart piano sonatas (36 distinct movements) in MIDI format by concert pianist Roland Batik, precisely aligned, at a note-by-note-level, to a standard edition (the New Mozart Edition) of the score, thereby linking the performance information to a previously published dataset \cite{hentschel2021annotated} of expert annotations of the scores in terms of harmony, cadence, and phrase structure.
To the best of our knowledge, this is the first corpus of its kind, combining high quality digital score and structural annotations with expert performances in recorded MIDI format. 
We report two preliminary experiments to demonstrate the benefits of having precise performance--score--structure annotation alignments.



\begin{table*}[ht]
    \centering

\small
\begin{tabular}{| l | S[table-format=4.0] | S[table-format=4.0] | c | c | c | c | c }

& \multicolumn{2}{|c|}{Size} & \multicolumn{2}{|c|}{Modality} & \multicolumn{2}{|c |}{Annotations} \\

Dataset & {Pieces} & {Performances} & MIDI & Score & Alignment & Other \\
\midrule
ASAP\cite{foscarin2020asap} & 222 & 1068 & recorded & MusicXML & beat & time and key signature \\
Vienna4x22\cite{vienna4x22} & 4 & 88 & recorded & MusicXML & note & - \\
CrestMuse PEDB\cite{hashida2018crestmusepedb} & 35 & 411 & recorded & MusicXML & note & phrase \\
MazurkaBL\cite{kosta2018mazurkabl} & 44 & 2000 & - & MusicXML &  beat & dynamics, tempo markings \\
\midrule
\midrule
\datasetname & 36 & 36 & recorded  & MusicXML & note & phrase, harmony, cadence
\end{tabular}

    \caption{An overview of publicly available comparable piano performance datasets for which precise recorded MIDI data, score-performance alignments and/or musicological annotations are available.}
    \label{tab:datasets_comparison}
\end{table*}

The remainder of this paper is organised as follows: Section \ref{sec:related} presents a list of comparable expressive performance datasets currently publicly available. Section \ref{sec:curation_and_formats} describes the data origins, the used data formats, and the curation process. Section \ref{sec:dataset_overview} gives an overview of the dataset, and Section \ref{sec:dataset_demo} describes two preliminary experiments to demonstrate the benefits of performance--score--structure annotation alignments. Finally, Section \ref{sec:conclusion} concludes the paper with some remarks for future work.

\section{Related Work}\label{sec:related}

Several piano performance datasets have been published in the context of expressive performance analysis and performance rendering. While recently published datasets are considerably larger than \datasetname, they provide performance recordings solely in the form of 
MIDI transcribed from audio recordings
\cite{zhang2022atepp, kong2020giantmidi} or do not include a high-quality digital score ground truth 
\cite{hawthorne2018enabling}. Despite the encouraging results demonstrated by recent transcription models, 
they often introduce inaccuracies, such as incorrect note fragmentation, missed note onsets, and falsely identified notes \cite{ycart2020investigating}. Similarly, certain expressive performance aspects such as (micro-)timing and tempo can only be measured given either a temporal or note-wise score-performance mapping \cite{cancino2018computational}. Nevertheless, these datasets remain useful for various related tasks such as symbolic music generation, music transcription and tagging, or 
high-level
comparative performance analysis.

Table \ref{tab:datasets_comparison} presents an overview of comparable piano performance datasets
currently publicly available, for which precise (recorded) MIDI data, score-performance alignments and/or musicological annotations are available.
Among these datasets, ASAP \cite{foscarin2020asap} stands out as the most extensive one, both in terms of musical pieces and performer range, with 1,068 performances beat-aligned to 222 scores, each annotated with key and time signature. In comparison to ASAP, all other publicly accessible datasets are significantly smaller: 
The Vienna 4x22 corpus \cite{vienna4x22} contains 22 different performances for excerpts of four different pieces, each aligned on a note level and provided in MusicXML\footnote{\url{https://www.musicxml.com/}}, MIDI and audio format. 
The CrestMuse PEDB v2.0 \cite{hashida2018crestmusepedb} provides 35 pieces note-aligned to 411 performances, with scores provided in MusicXML and MIDI and performances in MIDI and WAV. The dataset also contains phrase structure annotations, however, merely in the format of PDF and plain text files, somewhat limiting their (re)usability. 

The MazurkaBL dataset \cite{kosta2018mazurkabl} consists of a corpus of 44 Chopin Mazurkas with MusicXML scores that have been beat-aligned to 2000 performances. The performances themselves are not provided (neither as MIDI nor as audio); only beat positions and corresponding loudness values are given, along with the positions of tempo/dynamics markings in the score.



\section{Curation protocol and file formats}\label{sec:curation_and_formats}

\subsection{File origins}\label{sec:}

The MIDI performance files originate from a performance of twelve Mozart piano sonatas by
Viennese concert pianist
Roland Batik on a computer-controlled Bösendorfer SE290 grand piano, the predecessor of the CEUS model. The Bösendorfer SE series measures each individual keystroke and pedal movement precisely, with onset and offset times being captured at a time resolution of 1.25ms. Hammer velocity values are captured in a proprietary file format, and converted and mapped to the 128 dynamics MIDI values (see \cite{goebl2003measurement} for conversion details).
The audio recordings corresponding to those MIDI files can be purchased commercially\footnote{
\url{https://www.gramola.at/products/9003643987012}}.

These MIDI performance data were originally aligned manually, on a note-to-note level, to a symbolic encoding of the score produced by our team \cite{widmer2003discovering, cambouropoulos2000midi}. 
In order to make it possible to link the performance data in an unequivocal way to the musicological score annotations provided in the \textit{Annotated Mozart Sonatas} dataset by Hentschel et al. \cite{hentschel2021annotated}, we decided to replace our score encoding in the alignments entirely by the score notes as given in the their dataset, which link to their annotations directly via absolute temporal score position. 
The scores in the \textit{Annotated Mozart Sonatas} dataset conform to the New Mozart Edition\footnote{\url{https://dme.mozarteum.at/DME/nma/start.php?l=2}} and are given in MuseScore format, with the harmony, phrase and cadence label annotations provided in tabular format, as tab-separated values (TSV) files.

\subsection{The match alignment format}\label{matchfile_format}
We provide the alignment between the above-mentioned score and performance files in the match file format \cite{foscarin2022match}, a file format for symbolic music alignment in a human-understandable textual form. It is structured sequentially, and the alignment information is given at the level of individual notes.

The encoded alignment is complete in the sense that all performance and all score notes are captured. 
Each performance and each score note is represented with their respective note ID, 
and their respective alignment can be recorded with one out of three potential tuples: 1. A \textit{match} between score note and performance note, i.e., \texttt{(score\_id, performance\_id)}, 2. a \textit{deleted} score note \texttt{(score\_id, )} which represents a score note omitted in the performance, or 3. an \textit{inserted} performance note \texttt{( , performance\_id)}, which marks a performed note for which there is no corresponding score note.

Following this alignment encoding, each line in a match file corresponds to either a \textit{match}, a \textit{deletion} or an \textit{insertion}. Additional lines express (sustain or soft) pedal information, or encode meta information about the musical piece and performer. While the performance part in match corresponds to a lossless encoding of a corresponding performance in MIDI format, 
the score part captures essential information including onset, offset and duration in beats, and pitch, pitch spelling, and octave information for each score note.

\subsection{Curation protocol}\label{sec:curation_protocol}
To create note-level score-to-performance alignments, encoded in the match file format, between the performance MIDI data by pianist Roland Batik and scores and musicological annotations by Hentschel et al. \cite{hentschel2021annotated}, we follow the workflow as outlined below (see Fig. \ref{fig:mbb_alignment_process}):

\begin{figure}
 \centerline{\framebox{
 \includegraphics[width=0.9\columnwidth]{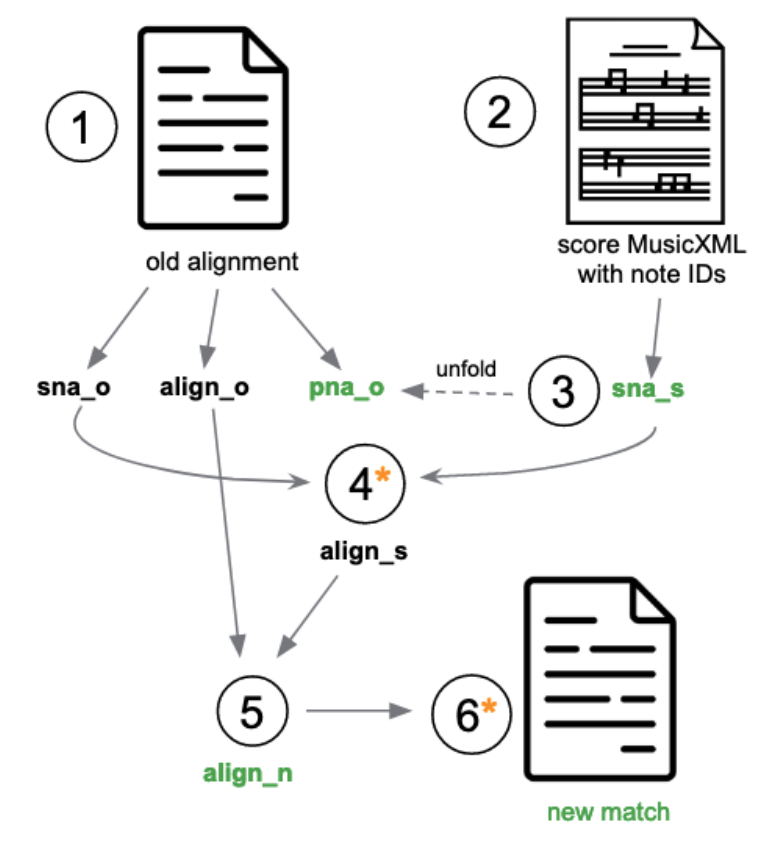}}}
 \caption{Visual illustration of the alignment process.
Each step in the alignment process is numbered according to the textual description in Section \ref{sec:curation_protocol}. Steps marked * indicate manual correction / post-processing. Elements highlighted in green are combined in the new alignment match files.}
\label{fig:mbb_alignment_process}
\end{figure}

\begin{enumerate}
    \item \textbf{Retrieve information from old alignment.} Given an old alignment file, we use partitura \cite{cancino2022partitura} to retrieve a score and performance representation which we parse into score and performance note arrays, \texttt{sna\_o} and \texttt{pna\_o}, to sequentially capture each (notated and performed) note with a unique note ID. In addition we retrieve a score-to-performance alignment, \texttt{align\_o}, in the encoding format explained above (i.e., a list of note ID tuples expressing either a match, deletion or insertion).
    
    \item \textbf{Retrieve score note array from MusicXML.} In the next step, we convert the annotated MuseScore format scores provided by Hentschel et al. \cite{hentschel2021annotated} to MusicXML, assign unique note IDs to each note, and convert this score representation into a second score note array (\texttt{sna\_s}).

    \item \textbf{Unfold score note array.} We update the score note array obtained from MusicXML, \texttt{sna\_s}, by unfolding it in accordance to the repetition structure found in the performance note array, \texttt{pna\_o}. \footnote{To reflect the same note occuring in a repeated segment, a suffix is added to the ID to reflect the number of occurrence, i.e. for a note with ID \texttt{n14}, the repeat structure unfolding is expressed as \texttt{n14-1} for the first, and \texttt{n14-2} for the second occurrence, respectively.}
    
    \item \textbf{Create score-score alignment.} In this step, we create a score-to-score alignment (\texttt{align\_s}) by matching each note in the two score note arrays \texttt{sna\_o} and \texttt{sna\_s} using its pitch, onset and duration information in beats.
    Any notes in \texttt{sna\_o} and \texttt{sna\_s} not matched automatically need to be aligned manually. 
    Missed alignments at this stage can occur due to:
    \begin{itemize}
        \item \textbf{Score mistakes}. These reflect mistakes in the score (e.g., a missing note, incorrect pitch, octave, missing modifier, missing repetition or ending markings) and require a manual correction of the score file.
        \item \textbf{Differing score versions}. For certain sonata movements, the notated score provides an alternative score version reflecting the first edition (``Erstdruck”) for certain segments of a piece, expressing the composer's impromptu ornamentation.\footnote{\url{https://www.henle.de/en/music-column/mozart-piano-sonatas/}} For the current dataset, such ornamented versions exist in K.284iii, K.332ii, K.457iii.
        \item \textbf{Double-voiced score notes}. These occur frequently in notated music, and describe a score note that is notated doubly in two different voices but corresponds to one performed note.
        \item \textbf{Grace notes}. Grace notes in notated music can occur in multiple forms to reflect different types of ornaments such as trills, acciaccature, mordents, turns etc. Depending on the ornament type and the underlying score encoding format, this may result in several notes occurring at the same (notated) onset (and hence with zero duration) to ensure a regular measure according to the time signature of that piece. Without onset and duration information, these notes must then be manually aligned to their corresponding performed notes.
        \item \textbf{Cadenza and \textit{ad libitum} measures.} Both cadenza measures and those marked \textit{ad libitum} correspond to irregular measures, that is, measures that contain more beats than indicated in the time signature (see Fig.~\ref{fig:cadenza_measure}). Digitally encoded, the notes in such measures are commonly notated without duration to allow for error-free parsing, and thus share the same beat onset and need to be aligned manually.
    \end{itemize}

    \begin{figure}[t]
     \centerline{
     \includegraphics[width=0.99\columnwidth]{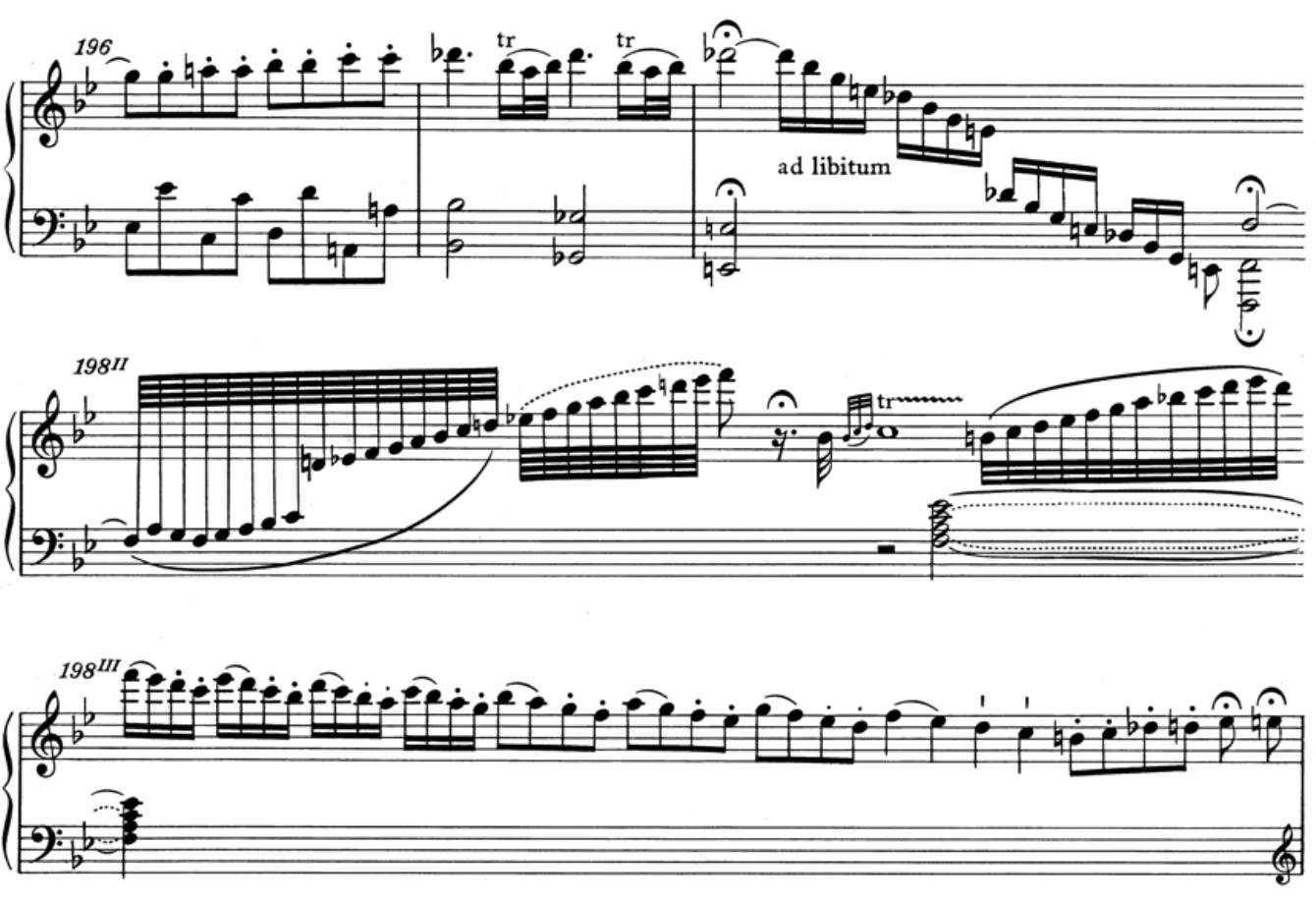}
     }
     \caption{An example of a cadenza within a piano sonata starting in measure 198 in KV333, 3rd movement.}
    \label{fig:cadenza_measure}
    \end{figure}
    
    \item \textbf{Update score-performance alignment.} Here we update the score note IDs in the old alignment (\texttt{align\_o}) according to the score-score alignment (\texttt{align\_s}) to create new score-performance alignments, \texttt{align\_n}.
    For each alignment in \texttt{align\_o}, we then need to ensure the validity of the original alignment type (match, insertion or deletion). In particular, for notes in the original score note array (\texttt{sna\_o}) that could not be aligned to notes in the MusicXML-based score note array (\texttt{sna\_s}), we consider two cases: 
    \begin{itemize}
        \item If the note in \texttt{sna\_o} corresponds to type ‘match’ in \texttt{align\_o}, the alignment type for the formerly matched performance note is changed accordingly into an insertion.
        \item If the note in \texttt{sna\_o} corresponds to type ‘deletion’ in \texttt{align\_o} (i.e., a score note that was not performed), it is is discarded in \texttt{align\_n}.
    \end{itemize}
    Notes in \texttt{sna\_s} that could not be aligned with notes in \texttt{sna\_o}, on the other hand, are recorded as type ‘deletion’ in \texttt{align\_n}. 
    \item \textbf{Create match files.} Using the updated performance-to-score alignment \texttt{align\_n}, we create new match files, and manually add attributional information (e.g., ‘diff\_score\_version’, ‘voice\_overlap’) to score notes to reflect edge cases described in step 4.
\end{enumerate}

\section{Dataset Overview}\label{sec:dataset_overview}

\begin{table*}[ht]
    \centering

\small
\begin{tabular}{
    |l
    S[table-format=4.0]
    S[table-format=2.3]
    |
    S[table-format=4.0]
    S[table-format=2.3]|
    S[table-format=4.0]
    S[table-format=2.3]|
    S[table-format=4.0]
    S[table-format=2.3]|
    }
\toprule
Sonata & {Performed Notes} & {Duration (min)} & {Match Notes} &  {\%} & {Insertion Notes} & {\%} & {Deletion Notes} &  {\%} \\ 
\midrule
 KV279 &        7789 &           16.21 &       7385 &     94.087 &          404 &        5.780 &       11 &    0.130 \\
 KV280 &        6277 &           14.69 &       6070 &     95.793 &          207 &        3.983 &       13 &    0.223 \\
 KV281 &        7030 &           14.43 &       6396 &     90.450 &          634 &        9.393 &       11 &    0.160 \\
 KV282 &        5761 &           14.77 &       5552 &     96.197 &          209 &        3.467 &       20 &    0.337 \\
 KV283 &        8231 &           17.39 &       7915 &     95.657 &          316 &        4.233 &        9 &    0.107 \\
 KV284 &       13386 &           25.92 &      12691 &     93.763 &          695 &        6.033 &       27 &    0.203 \\
 KV330 &        7869 &           18.47 &       7589 &     96.857 &          280 &        3.047 &        7 &    0.100 \\
 KV331 &       11760 &           22.64 &      11595 &     98.283 &          165 &        1.370 &       45 &    0.347 \\
 KV332 &        9013 &           17.84 &       8660 &     93.417 &          353 &        6.210 &       24 &    0.373 \\
 KV333 &        9137 &           20.40 &       8827 &     96.723 &          310 &        3.120 &       16 &    0.157 \\
 KV457 &        7290 &           18.24 &       7022 &     96.043 &          268 &        3.843 &        9 &    0.110 \\
 KV533 &        8878 &           22.12 &       8616 &     97.027 &          262 &        2.837 &       15 &    0.137 \\
\midrule
 Total &      102421 &          223.12 &      98318 &     95.358 &         4103 &        4.443 &      207 &    0.199 \\
\bottomrule
\end{tabular}

    \caption{List of sonatas in the \datasetname dataset. The bottom row represents the sum in all columns except for those expressing percentages, for which the mean is shown.}
    \label{tab:mozart_batik_corpus}
\end{table*}

The \datasetname dataset contains performances by pianist Roland Batik
of twelve Mozart sonatas (see Table \ref{tab:mozart_batik_corpus} for the list of sonatas), corresponding to approx.~102,400 played notes and 223 minutes of music, for which the performances are provided in MIDI, musical scores in MusicXML, and the alignment in match file format. Approximately 98,300 (95.36\%) of all performed notes are aligned with a corresponding score note, the remaining 4,100 (4.44\%) represent insertions (reflecting mostly ornaments). Roughly 200 score notes have been omitted in the performances.

For each performance, we also provide the performance note arrays, which capture each played note with its note ID along with onset and duration information in seconds and MIDI ticks, as well as velocity and pitch information.
Likewise, the dataset includes the score note array (unfolded according to the repeats as played by the pianist and reflected in the alignment), which captures each score note with its (MusicXML) note ID (including repeat suffices, where applicable), onset and duration information in terms of beats (reflecting the time signature), and quarter notes (reflecting a “normalized” score time unit), and pitch and voice information. 

We link our aligned score note arrays to the musicological annotations in \cite{hentschel2021annotated} via their temporal position in the following way: 
In the second version\footnote{
\url{https://github.com/DCMLab/mozart\_piano\_sonatas}} of the dataset, each annotation label for harmonies, cadences, and phrases is unequivocally referenced to a temporal score position represented in terms of quarterbeats and measure number, where the first expresses the distance of the label from the beginning of the piece in quarter note units. We leverage these two temporal parameters to link each note-aligned score note array by first reducing it to its shortest form (without any unfolded repeats), aligning it temporally with the musicological annotations, and eventually unfolding it according to the performed repetition structure. 

\section{Dataset Demonstrations}\label{sec:dataset_demo}

This section presents two simple examples of the kinds of studies that are made possible by our dataset. The first is motivated by a directly related study in the Annotated Mozart Sonata corpus paper \cite{hentschel2021annotated}; the second shows how precise performance alignments permit more detailed investigations relating to cadences and their performance.

\subsection{Global tempo and harmonic density}

In a first study, we replicate the second experiment in Hentschel et al. \cite{hentschel2021annotated}, aimed at investigating the relationship between tempo and harmonic change rate. The basic question asked in \cite{hentschel2021annotated} was whether the rate at which the harmony changes in a piece is correlated with the piece's typical performance tempo. Their study involved determining the average (median) performance duration of each sonata movement from 6 complete commercial sonata recordings, and correlating harmonic label density (rate of harmonic labels in their annotations, per performance time unit) with average overall performance tempo (number of quarter notes per performance time unit).
We repeat the same experiment with our pianist's performances and our alignment files instead of 6 pianists' audio recordings.

We apply the same procedure as in \cite{hentschel2021annotated}, unfolding the score according to the repeat structure of the piece in order to calculate the actual piece length (in terms of quarter notes). The only difference is that we do this according to the repeats actually performed by the pianist (which are expressed in our match files, thus omitting the need for a dedicated ``unfolding" step), whereas \cite{hentschel2021annotated} seem to have assumed that all repeats were played by all pianists.

\begin{figure}[t]
 \centerline{
 \includegraphics[width=1.0\columnwidth]{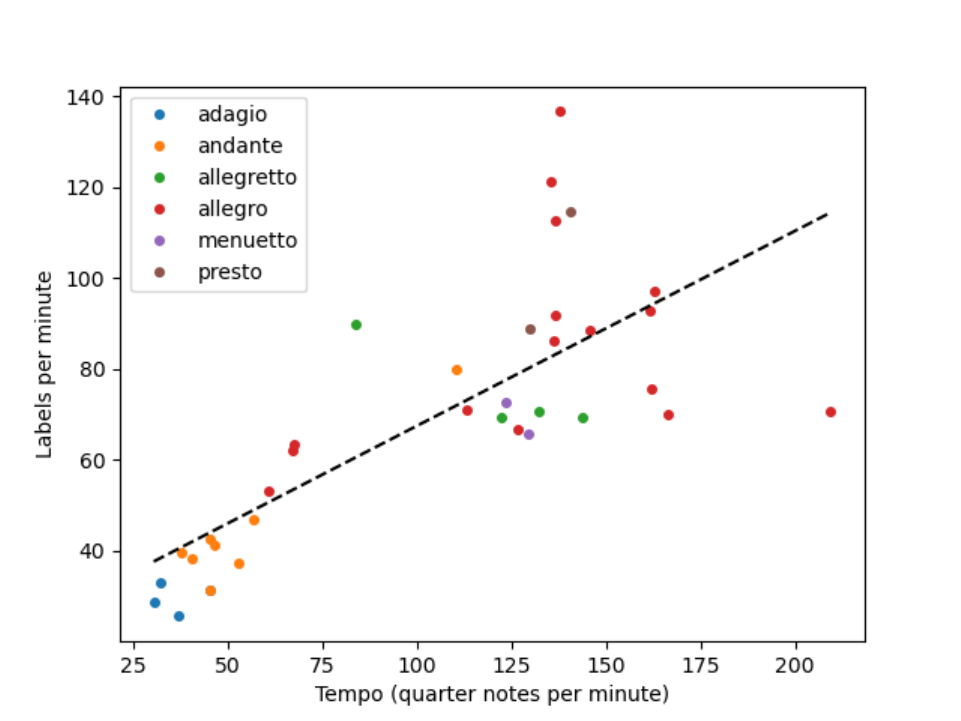}
 }
 \caption{Correlation between global tempo (as measured in quarter notes per minute) and harmony label density}
\label{fig:exp1_recplication}
\end{figure}

Comparing our results (Fig.~\ref{fig:exp1_recplication}) to Fig.~\href{https://dcmlab.github.io/mozart_piano_sonatas/10.html}{10} in \cite{hentschel2021annotated}, we see a similar general trend, in the form of a roughly linear increase in harmonic label
density with performed tempo (slope = .43, r= .75, compared to .48 and .80,
respectively, in \cite{hentschel2021annotated}).\footnote{Note that we have a somewhat smaller
set of points, because we only have 12 of the 18 sonatas in our dataset.}
However, we also immediately see a marked difference in the performance tempo
distribution: in \cite{hentschel2021annotated}, Fig. \href{https://dcmlab.github.io/mozart_piano_sonatas/10.html}{10}, there is a relatively large cloud of points (sonata movements)
with conspicuously high tempos of 180--200 (quarters per minute), which
does not appear in our plot, and which we believe may point to a systematic
problem in their way of estimating playing tempo: assuming that
all notated repeats are played out by the performers leads them to
overestimate the tempo in all cases where some or a majority skipped some repeats.\footnote{
Of course, the authors explicitly acknowledge the problem:
``Also, some of the initial assumptions might have to be revisited.
For example, the extreme outlier suggesting a tempo of 239 quarter notes per minute
is due to the fact that for this particular piece -- the first movement of K. 533/494 -- there seems to be a convention among pianists to repeat the first part of the piece, but not the second (as the score would suggest), which of course reduces the performance duration.'' \cite{hentschel2021annotated} (p.76), but a comparison with our distribution implies it might be more severe than expected.}

We thus see an immediate advantage of our more precise performance-aligned corpus:
the match files naturally give correct tempo and score duration
information, being based as they are on score-performance alignments
that reflect the actual repeat structure played by our performer.
Still, we can say that our results support and confirm the overall hypotheses proposed there,
showing a more or less linear relationship between harmonic label density
and global performance tempo.

\subsection{Performance of different cadence types}

\begin{figure*}[ht]
 \centerline{
 \includegraphics[width=.92\textwidth]{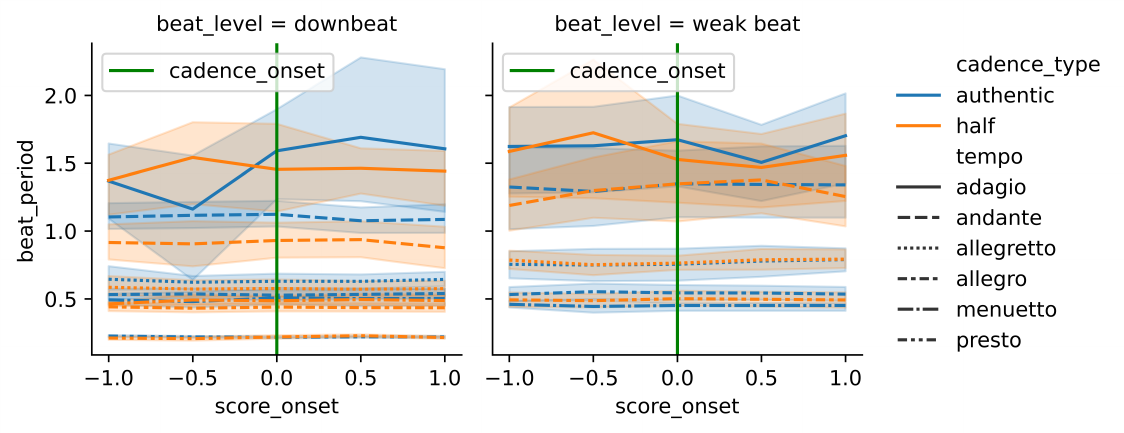}
 }
 \caption{
 Comparison of local timing strategies one quarter note before and after authentic and half cadence labels, over different tempo classes (in increasing tempo from top to bottom), for cadences falling on a downbeat (left) or weak beat (right).
 Colour identifies cadence type, line style notated tempo class.
 }
\label{fig:cadence_tempo}
\end{figure*}

Our data permits much more detailed investigations into relationships between structural aspects of a piece, and how these are translated into performance decisions by a pianist. As a simple example, we investigate variations in local tempo before various types of cadences. Specifically, we compare the local tempo prior to a cadence annotation across different tempo classes for authentic (perfect and imperfect, i.e., PAC and IAC) and half cadences (HC), and differentiate between the cases when a cadence falls on either a downbeat or a weak beat. 
The hypothesis to be tested here is that a performer will tend to shape cadences differently, in terms of tempo, depending on their type and degree of `finality'.

To compute the local tempo curves, we consider a uniform window spanning one quarter note each preceding and following a cadence label.\footnote{
In the \textit{Annotated Mozart Sonatas Corpus} \cite{hentschel2021annotated}, cadence labels are placed at the onset of the final target harmony (e.g., I/i for authentic cadences).}
For each score-note-aligned performed note in that window, we define the local tempo via the \textit{beat period (BP)}, which we calculate as the ratio of the inter-onset-interval (IOI) between the current \textit{performed} onset and the subsequent one, and the IOI between the current \textit{notated} onset and subsequent one. We exclude grace notes and their corresponding performed notes from this calculation in order to remove outliers.

Next, we perform time-wise interpolation on these tempo curves to obtain beat period values at eighth note intervals within the window. Given that we are most interested in the local timing strategy immediately before a label (that is, an eighth note before the label position), we discard those curves where that particular time point is interpolated. Following this procedure, we obtain a total of 3,540 local tempo values (corresponding to 708 curves), of which 251 (7.09\%) values are interpolated.

Figure \ref{fig:cadence_tempo} shows the mean of local tempo curves across different tempo classes, for cadence labels annotated on a downbeat (left) and on a weak beat (right), respectively. For both authentic and half cadence types, the differences in local tempo diminish with increasing global tempo for both downbeat and weak beat cadences.
Likewise, the tempo profiles tend to flatten out with increasing global tempo, suggesting that the pianist takes more liberty, in terms of expressive timing, in slow pieces.
For this reason, we focus our analysis on the \textit{adagio} tempo class, the slowest tempo 
(the solid line plots in Fig.~\ref{fig:cadence_tempo})
.

The influence of the beat level on the local tempo for half cadences seems to be negligible, with the local beat period decreasing slightly prior to the cadence (causing an increase in local tempo, i.e. an \textit{accelerando}), regardless of whether it falls on a downbeat or weak beat. 
For authentic cadences, we can see a substantial difference in expressive tempo depending on whether or not the label falls on a downbeat: for authentic cadences falling on a downbeat, the mean tempo curve for the \textit{adagio} tempo class corresponds mostly to what one would expect (i.e., a very clear \textit{ritardando} in preparation of the cadence) based on the underlying harmonies and their notion of tension and release. Interestingly, this \textit{ritard} seems to continue somewhat after the resolution into the tonic, suggesting a lengthening of the tonic arrival. For weak-beat authentic cadences, a similar significant preparation or anticipation is largely missing.

\section{Conclusion and Future Work}\label{sec:conclusion}

We have presented \datasetname, a piano performance dataset linking professional Mozart piano sonata performances to expert-labelled musical scores, at the level of notes. The resulting dataset is the first of its kind to combine professional performances in precise, recorded MIDI with curated musical scores and expert musicological and structural annotations \cite{hentschel2021annotated} at this level of detail. 

We presented two preliminary experiments, intended to demonstrate the benefits of having such precise, note-aligned performance--score--structure annotation data for studying expressive features and their relation to the underlying musical structure.

Our plan for future work includes the transcription of the remaining six sonatas of the Mozart piano sonatas corpus from audio recordings by the same pianist,
and their subsequent alignment to the musical scores using state-of-the-art transcription and alignment models. By doing so, we hope to advance our understanding of the differences between transcribed and recorded MIDI, and to evaluate the potential benefits of incorporating an alignment step to improve the quality of transcription.


\section{Acknowledgments}
We wish to express our gratitude to pianist Roland Batik for his gracious permission to publish the detailed measurements of his performances. We also want to thank the authors of the \textit{Annotated Mozart Sonatas Corpus} for their tremendous efforts, and for permitting us to link our data to theirs.
This work receives funding from the European Research Council (ERC), under the European Union's Horizon 2020 research and innovation programme, grant agreement No.~101019375 (\textit{Whither Music?}). The LIT AI Lab is supported by the Federal State of Upper Austria.

\bibliography{ISMIR2023_template}



\end{document}